# Oscillations and damping in the fractional Maxwell materials

Robyn H. Pritchard and Eugene M. Terentjev[a)]

*Cavendish Laboratory, University of Cambridge, JJ Thomson Avenue, Cambridge CB3 0HE, United Kingdom*

## Abstract

This paper examines the oscillatory behavior of complex viscoelastic systems with power law like relaxation behavior. Specifically, we use the fractional Maxwell model, consisting of a spring and fractional dashpot in series, which produces a power-law creep behavior and a relaxation law following the Mittag-Leffler function. The fractional dashpot is characterized by a parameter $\beta$, continuously moving from the pure viscous behavior when $\beta = 1$ to the purely elastic response when $\beta = 0$. In this work, we study the general response function and focus on the oscillatory behavior of a fractional Maxwell system in four regimes: Stress impulse, strain impulse, step stress, and driven oscillations. The solutions are presented in a format analogous to the classical oscillator, showing how the fractional nature of relaxation changes the long-time equilibrium behavior and the short-time transient solutions. We specifically test the critical damping conditions in the fractional regime, since these have a particular relevance in biomechanics. © *2017 Author(s). All article content, except where otherwise noted, is licensed under a Creative Commons Attribution (CC BY) license (http://creativecommons.org/licenses/by/4.0/).* [http://dx.doi.org/10.1122/1.4973957]

## I. INTRODUCTION

The damping of free oscillations is a subject with a long and rich history. However, there is less work available looking at how the damping of oscillations depends specifically on the viscoelasticity of a system, and in particular, the types of viscoelasticity seen in biological tissues and complex composite materials. The relaxation function in many such systems is distinctly nonexponential, which often referred to as a distribution or spectrum of internal relaxation times. An alternative way to describe such complex viscoelasticity is via a viscoelastic element characterized by a fractional time derivative. In this work, we examine the oscillatory behavior of a viscoelastic element characterized by a fractional Maxwell model [1–3]. The fractional Maxwell model is a generalization of the classical model (based on the elastic and damping elements in series) and successfully describes the relaxation function of polymeric materials [4–7], biological tissues [8–10], cells [11,12], and foods [13–15] over a large range of time scales. It is readily generalized to a fractional Zener model [16,17] for the description of viscoelastic solids (since the Maxwell model ultimately produces a plastic flow in response to a constant force). It is useful to be able to relate the well-understood characteristics of viscoelastic behavior to how a system dissipates energy in free oscillations. This in particular can give insight into the role biological tissues have in dissipating energy from their relaxation function.

This work draws from the rich field of fractional oscillators, with earlier rigorous work of Rossikhin and Shitikova looking at its applications in viscoelasticity [18,19]. Fractional dynamics involves taking the integer order differential operators and replacing them with operators of a fractional order [20,21]; however, in many papers which use fractional dynamics there is no direct physical justification for this. This is particularly true for harmonic oscillators where the inertial term is taken as fractional (with arguably unclear physical foundations). Here, we retain a standard inertia effects, but include the fractional-Maxwell relaxation function that models realistic viscoelasticity, and use it as the origin of elasticity and damping to derive oscillatory response as a result to several representative initial conditions.

In this paper, we first review the linear viscoelasticity and the corresponding relaxation function to establish a reference point, and then justify the use of the fractional Maxwell model and where it might apply. We then derive the general response function of a viscoelastic element characterized by a given relaxation function, $G(t)$. Sections III A–C discuss the response of the fractional Maxwell model to a stress impulse, strain impulse, step stress, and driven oscillations, while drawing an analogy to the classical case. Sections IV and V discuss the frequency and decay of free oscillations for the fractional Maxwell model. The interest here is to give insight into the damping of oscillations in biological materials, which typically have a more complicated viscoelasticity of a fractional type. One example would be a tendon, an essential mechanical element of large organisms, which should have low damping and respond in a similar way to a range of frequencies of oscillation to fulfill its biological function. On the other hand, for other soft connective tissues it would be preferable to be positioned close to critical damping conditions to prevent injury by efficiently dissipating any excess mechanical energy from impacts that could not be fully absorbed by muscle (such as in running or landing from a jump).

## II. VISCOELASTIC MODELS

In this section, we briefly review basic viscoelasticity and examine relaxation functions. We also demonstrate the

---

[a)]Author to whom correspondence should be addressed; electronic mail: emt1000@cam.ac.uk

crossover between some of the phenomenological models and how often it can be difficult to distinguish among them. In linear viscoelasticity, the stress $\sigma$ arising in response to some arbitrary strain, $\varepsilon(t)$, is determined by the Boltzmann superposition integral [22]

$$\sigma(t) = \int_{-\infty}^{t} G(t - t')\dot{\varepsilon}(t')\,dt', \qquad (1)$$

where $G(t)$ is the characteristic linear response function of the material, often called the relaxation function. We can see that $G(t)$ is measured by the system response to a step strain, where $\dot{\varepsilon}(t) = \Delta\varepsilon \cdot \delta(t')$ and Eq. (1) becomes $\sigma(t) = G(t) \cdot \Delta\varepsilon$. The Boltzmann superposition integral itself is the limit sum over infinitesimal step strains determined by $d\varepsilon = (d\varepsilon/d\tau)d\tau$. Figure 1(a) shows a schematic of typical relaxation curve, which includes the key features of: (a) The instantaneous response to step strain is often called the glass modulus; (b) the modulus relaxes over some characteristic time, $\tau$, to an equilibrium modulus $G_e$; (c) the amount of modulus relaxation is labeled as $G_r$.

The relaxation time $\tau$ is only clearly defined in the case of simple exponential relaxation $G(t) = G_e + G_r e^{-t/\tau}$. However, it is very rare that a practical material can be described with a single exponential relaxation, except in the cases of transient networks with a single rate of crosslink breaking [23]. The attempts to rationalize a more complex relaxation response often apply a "brute force" fit with multiple relaxation times, the so called generalized Maxwell model. Another common relaxation function that fits well with materials with a fractal or hierarchical structure is the power law. This is often called scale-free rheology as there is no end to relaxation that follows $At^{-\beta}$ at all probed time scales, with only the magnitude $A$ changing. An issue with power law relaxation is that there is no finite modulus at $t = 0$, an issue usually avoided by using the asymptotic power law $1/(1 + At^{\beta})$.

Viscoelasticity is often described through phenomenological models built of springs and dashpot elements of the basic Maxwell model, the simplest alternatives being the Kelvin–Voigt model where the spring and dashpot are connected in parallel and the Zener model where there is one additional elastic element in series with the Voigt frame. However, to fit realistic relaxation functions a great many dashpots and springs are often required, and with the number of parameters increasing the basic physical sense rapidly dissipates. The ideas of fractional viscoelasticity have been developed to address the power-law relaxation observations while retaining a fixed small number of fitting parameters. In the words of [2], it achieves a "mimicry of memory." In this case, the dashpots are replaced with fractional dashpots (sometimes called Scott-Blair elements), which on their own follow power-law relaxation [24]. The use of fractional dashpot changes the constitutive equation from $\sigma = \eta\,d\varepsilon/dt$ to $\sigma = \eta^{\beta}d^{\beta}\varepsilon/dt^{\beta}$, where the fractional derivative is a formal generalization of a derivative to noninteger order (which in practice is only meaningful in the reciprocal representation of Fourier or Laplace transformations). Sometimes, the fractional Maxwell model is presented as two fractional dashpots in series, which gives an extra fitting parameter; however, for simplicity we focus on the case with just one fractional dashpot. For a good introduction to fractional derivatives in viscoelasticity, see [3,25], while the general matters of fractional calculus can be reviewed in [26].

Table I shows several model viscoelastic relaxation functions and their Laplace transforms. The fractional Maxwell model is useful to model a variety of relaxation processes by adding only one additional parameter, $\beta$. This is opposed to more complicated models, such as multirelaxation models which may need a great many parameters before a good fit is found. It also differs from other relaxation functions, such as the stretched-exponential and nonasymptotic power law, which do not have phenomenological models to underpin the empirical expressions. The step strain response of an isolated fractional dashpot element is simply a power law of the form $At^{-\beta}/\Gamma[1 - \beta]$, where $\Gamma$ is gamma function. When $\beta = 0$ the fractional dashpot acts as a perfectly elastic solid; as $\beta$ increases, the element becomes more viscous, until the point of $\beta = 1$ is reached, when $(At^{-1}/\Gamma[0]) = A\delta(t)$ becomes the relaxation of a pure Newtonian fluid.

The fractional Maxwell model has its relaxation described by the Mittag-Leffler function, $E_{\beta}[-(t/\tau)^{\beta}]$ (see Appendix A for details), which behaves very similarly to the nonasymptotic power law, $1/(1 + (t/\tau)^{\beta})$, being equal to one at $t = 0$ and converging on the power law at long times. For small $\beta$, the Mittag-Leffler function is fitted exceptionally well by the asymptotic power law, which is also roughly equivalent to the stretched exponential with $1/2\exp[1 - (t/\tau)^{\beta/2}]$. In many situations, these functions are equivalent, or practically indistinguishable over typical experimental time scales [27], as shown in Fig. 2.

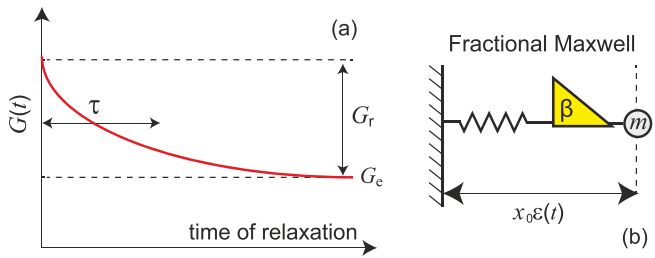

FIG. 1. (a) Scheme of a typical stress relaxation curve, where $G_r + G_e$ is the instantaneous (glass) response to a step strain. (b) Scheme of the Maxwell model with the two elements in series, with an inertial element (mass) attached to one end. When $\beta = 1$, the fractional dashpot becomes a regular damping element of the classical Maxwell model. The variable $x_0\varepsilon(t)$ is the distance of the mass from the wall, where $x_0$ is the rest length of the system.

TABLE I. An example of several relaxation (response) functions and their corresponding Laplace transforms. $E_{\beta}[z]$ is the Mittag-Leffler function of rank $\beta$, see text.

| Model | Response, $G(t)$ (Pa) | $\tilde{G}(s)$ (Pa s) |
|---|---|---|
| Kelvin–Voigt | $\eta\delta(t) + G_e$ | $\eta + \dfrac{G_e}{s}$ |
| Maxwell | $G_r \exp(-t/\tau)$ | $G_r\,\dfrac{1}{s + 1/\tau}$ |
| Fractional Maxwell | $G_r E_{\beta}[-(t/\tau)^{\beta}]$ | $G_r\,\dfrac{s^{\beta-1}}{s^{\beta} + (1/\tau)^{\beta}}$ |
| Fractional dashpot | $\dfrac{A}{\Gamma[1 - \beta]}t^{-\beta}$ | $A s^{\beta-1}$ |

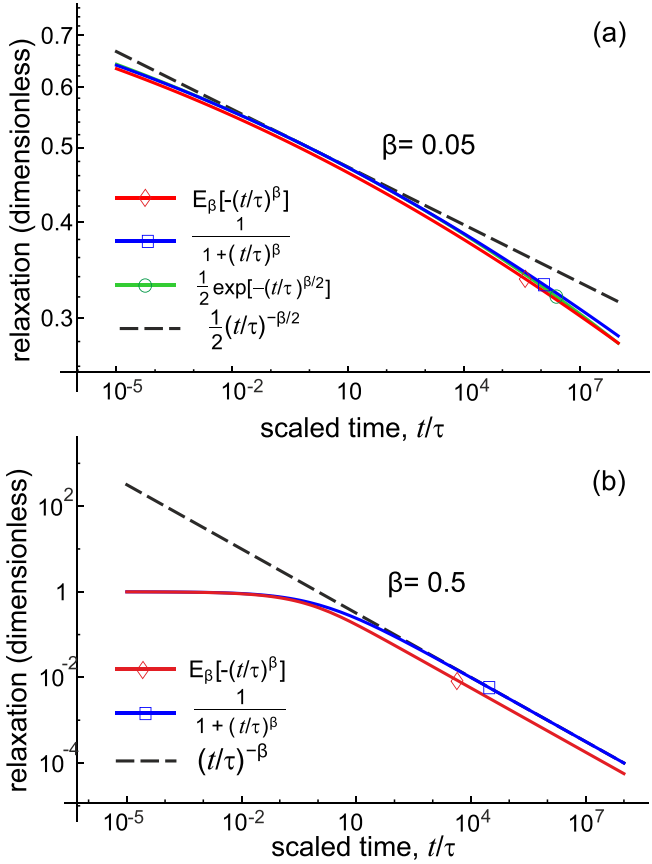

**FIG. 2.** (a) The Mittag-Leffler function for small $\beta$ compared with traditionally used relaxation functions, illustrating how difficult it could be to distinguish between them in practice. (b) The Mittag-Leffler function for a much higher $\beta = 0.5$ compared to the nonasymptotic power and power law, again highlighting the similarity at long times.

It is clear that in a great many situations the Mittag-Leffler function is as good a fit, if not better, than the power-law, asymptotic power law, or a stretched exponential. It has the key benefits of a finite value at zero time, having few parameters, and having a compact Laplace transform which is useful for any analytical calculations. The Mittag-Leffler function is frequently used to fit the relaxation in polymeric materials [28]. In particular, in biological tissues [29–31] and cells [32–34], the stress relaxation often exhibits low power laws, which lend themselves being fit to the Mittag-Leffler function. For example, the Mittag-Leffler function has been used to fit human breast tissue cells relaxation function with $\beta \approx 0.5$ [35].

The following sections examine the oscillatory response of the fractional Maxwell model in response to an impulse, impulses strain, step strain, and driven oscillations. In each case, we first solve the classical Maxwell model to highlight the differences and similarities.

## III. GENERAL SOLUTION

Let us examine the simple model of a mass attached to a generic viscoelastic element, with the other end fixed, and derive the general solution for its motion, see Fig. 1(b). The position of the mass is governed by its inertia, an external force it is subjected to, $F(t)$, and the relaxation function of the viscoelastic material, $G(t)$. Thus, the equation of motion takes the form

$$m\frac{\mathrm{d}^2 x(t)}{\mathrm{d}t^2} = -A \int_{-\infty}^{t} G(t-\tau)\frac{\mathrm{d}\varepsilon}{\mathrm{d}\tau}\mathrm{d}\tau + F(t), \quad (2)$$

where $x = x_0\varepsilon(t)$ is the position of the mass, $A$ is the cross-section area, and $\varepsilon$ is the strain in the viscoelastic element. This is a balance between inertial forces, the stress in the viscoelastic element given through the Boltzmann superposition integral, and the external force. As $\mathrm{d}x/\mathrm{d}t = x_0\mathrm{d}\varepsilon/\mathrm{d}t$, Eq. (2) can be written in terms of stress and strain

$$\frac{\mathrm{d}^2\varepsilon(t)}{\mathrm{d}t^2} = -\frac{1}{\mu}\int_{-\infty}^{t} G(t-\tau)\frac{\mathrm{d}\varepsilon}{\mathrm{d}\tau}\mathrm{d}\tau + \frac{1}{\mu}\sigma(t), \quad (3)$$

where $\mu = mx_0/A$ is the reduced mass and $\sigma(t) = F(t)/A$ is the external stress, respectively. The Laplace transformation of Eq. (3) gives a general relation

$$s^2\tilde{\varepsilon}(s) - s\varepsilon(0) - \dot{\varepsilon}(0) + \frac{\tilde{G}(s)}{\mu}\left(s\tilde{\varepsilon}(s) - \varepsilon(0)\right) = \frac{\tilde{\sigma}(s)}{\mu}. \quad (4)$$

We assume that until $t = 0$ the mass was stationary, such that $\varepsilon(t) = 0$ and $\dot{\varepsilon}(t) = 0$ (no infinite acceleration is allowed). The nontrivial initial condition that initiates motion is that at $t = 0$ a stress $\sigma(t)$ is applied. Under these conditions, we can write the general solution in Laplace space

$$\tilde{\varepsilon}(s) = \frac{\tilde{\sigma}(s)}{\mu s^2 + s\tilde{G}(s)}. \quad (5)$$

To find the strain response as a function of time, we must first substitute into Eq. (5) the Laplace transform of the external stress, which reflects the specific mode of dynamic experiment, and also the specific viscoelastic relaxation function, which reflects the physical nature of the material. After this, the inverse Laplace transformation has to be performed. Table II shows the four characteristic experiments, which produce initial conditions on the externally applied stress that we look at in this paper, along with their Laplace transforms. The characteristic of an impact is a "stress impulse" (this solution is also called the transfer function). In this case, the external condition is a step-function change in momentum (or velocity) centered at $t = 0$, which manifests itself as a delta-function stress. A step stress is equivalent to having a weight suddenly attached, allowing the system to find a new equilibrium position. The strain impulse is the act of "plucking" the mass, that is, moving it instantaneously from zero strain to some finite strain and then releasing: The stress function used in this initial condition may seem strange, but the same general solution in Laplace space can be found by having no initial stress and setting $\varepsilon(0)$ to $\Delta\varepsilon$. In this case, an instantaneous change in strain is reflected by a step-function change in position at $t = 0$, which corresponds to the derivative of the delta function in force. The final initial condition is the response to driven oscillations, which is another focus of this paper.

**TABLE II.** Four external stress conditions and their corresponding Laplace transforms. The first three cases arise from a discontinuity that occurs between the ambient state before $t = 0$ and the solutions for $t > 0$.

| Initial condition | $\sigma(t)$ (Pa) | $\tilde{\sigma}(s)$ (Pa s) |
|---|---|---|
| Impulse stress | $\Sigma_0\delta(t)$ | $\Sigma_0$ |
| Step stress | $\sigma_0$ | $\sigma_0/s$ |
| Impulse strain | $\mu\Delta\varepsilon(\mathrm{d}\delta(t)/\mathrm{d}t)$ | $\mu\Delta\varepsilon$ |
| Driven oscillation | $\sigma_0\sin(\omega t)$ | $\sigma_0\omega/(\omega^2 + s^2)$ |

In each case, before solving for the fractional Maxwell model for each case of the initial condition, we first solve for the classical Voigt and Maxwell models. The Voigt model has the same solutions as the linear damped oscillator, which we use to remind the reader of the familiar classical results. The solution of the classical Maxwell model ($\beta = 1$) allows to draw a useful analogy and highlight the origin of the different terms in the fractional Maxwell solutions. For clarity, Table III lists the general solutions in Laplace space, for the initial conditions and models looked at in this paper. The expressions in this table use shorthand notations, which are important to list here to avoid ambiguity, as they will be frequently used below. In the Voigt model, $\omega^2 = \omega_{eq}^2 - \lambda^2$, with $\lambda = \eta/2\mu$, and $\omega_{eq}^2 = G_e/\mu$. In the Maxwell models, the commonly used parameters take the form: $\omega^2 = \omega_\infty^2 - \lambda^2$, with $\lambda = 1/2\tau = G_r/2\eta$, and $\omega_\infty^2 = G_r/\mu$.

## A. Response to a stress impulse

In this case, the momentum is instantaneously changed at $t = 0$ from 0 to some finite value, i.e., from stationary the mass is given an initial velocity via impact. The initial stress condition can be written as $\sigma(t) = \Sigma_0\delta(t)$ where $\Sigma_0$ has units of $\mathrm{kg\,m^{-1}\,s^{-1}}$. The equation of motion is Laplace space is then

$$\bar{\varepsilon}(s) = \frac{\Sigma_0}{\mu s^2 + s\bar{G}(s)}. \quad (6)$$

The classical case of damped oscillations is equivalent to using the Voigt model, where an elastic restoring force and a viscous damping force act in parallel (viscous dissipation could either be due to internal or external friction). Taking the inverse Laplace transform of the solution in Table III gives the three modes of response depending on whether $\omega$ is zero, real, or imaginary. These modes are shown in Table IV.

These three solutions correspond to the classical cases of under-damped, over-damped, and critically damped oscillation, respectively. The response in this case of a Voigt model, as expected, settles to an equilibrium at zero strain, and has no long-term memory of it being subjected to an impulse.

Now looking at the Maxwell model, and taking the inverse Laplace transformation of the solution listed in Table III, we obtain

$$\varepsilon(t) = \frac{2\lambda\Sigma_0}{\mu\omega_\infty^2}\left(1 - \frac{\omega_\infty^2}{2\lambda\omega}\cos(\omega t + \phi)\exp(-\lambda t)\right), \quad (7)$$

see Appendix B for derivation. At $t = 0$, the terms in the bracket cancel exactly, as we would expect. At long times, the oscillatory part decays and the strain settles at the equilibrium position $\varepsilon_\infty = \Sigma_0/\eta$ [after simplifying the prefactor in Eq. (7)]. This is the unrecoverable deformation that happened through the dashpot creep in a Maxwell model. As with the Voigt model, we can identify three distinct regimes of underdamped oscillations for $\omega_\infty > \lambda$, slow overdamped relaxation for $\omega_\infty < \lambda$, and the fastest approach to equilibrium for $\omega_\infty = \lambda$ (critical damping).

Now turning to the fractional Maxwell model, taking the inverse Laplace transformation of the corresponding solution in Table III is more difficult due to noninteger powers; the full derivation is discussed in Appendix C. Once solved, the response of this viscoelastic system to a stress impulse takes the form

$$\varepsilon(t) = \frac{\Sigma_0}{\mu t\omega_\infty^2}\left(\frac{t^{\beta-1}}{\tau^{\beta-1}\Gamma[\beta]} - A\cos[\omega t + \phi]\exp(-\lambda t)\right.$$
$$\left. - \frac{\sin[\pi\beta]}{\pi}\int_0^\infty \rho(x)x^{-\beta}\exp(-xt/\tau)\mathrm{d}x\right), \quad (8)$$

where the values of shorthand parameters $A$, $\phi$, and the function $\rho(x)$ are listed in Appendix C. The values of $\omega$ and $\lambda$ are determined by the roots of $s^2 + \tau^{-1}s^{2-\beta} + \omega_\infty^2 = 0$, because for the inverse Laplace transformation of a function with simple poles, the only time dependent part is the exponential, i.e., the locations of the poles directly determine the frequency and decay rate of the oscillations. When $\beta = 1$, we recover the classical Maxwell model. The integral term vanishes quickly and is mainly responsible for stabilizing the behavior at short times, in particular the asymptotic behavior of the power law as $t \to 0$. Equation (8) actually proves a conjecture made in [36], referring to a criterion for the crossover between liquid-and solidlike behavior of a viscoelastic system.

There is a small technical point to be made here. In general, the integral with $\rho(x)$ in Eq. (8) cannot be evaluated analytically. In most cases, it converges well and can easily be found numerically, and the dependence on key parameters extracted explicitly. However, in some cases (usually when $\omega\tau \ll 1$) the convergence is poor and the result could be less clear. To address this issue, here and in similar expressions below, we deliberately isolate the integral which is going to play as small a role as possible, bringing out the explicit long-time contributions which can be solved exactly.

TABLE III. General solutions for the strain in Laplace space, $\bar{\varepsilon}(s)$, obtained from Eq. (5) and Tables I and II. The notations for $\omega$, $\omega_\infty$, and $\lambda$ in each case are detailed in the text.

| Model | $\bar{\varepsilon}(s)$: Impulse stress | $\bar{\varepsilon}(s)$: Impulse strain | $\bar{\varepsilon}(s)$: Step stress |
|---|---|---|---|
| Kelvin–Voigt | $\dfrac{\Sigma_0}{\mu\omega}\dfrac{\omega}{(s+\lambda)^2+\omega^2}$ | $\Delta\varepsilon\dfrac{s}{(s+\lambda)^2+\omega^2}$ | $\dfrac{\sigma_0}{\mu}\dfrac{s^{-1}}{(s+\lambda)^2+\omega^2}$ |
| Maxwell | $\dfrac{\Sigma_0}{\mu}\dfrac{1+2\lambda s^{-1}}{(s+\lambda)^2+\omega^2}$ | $\Delta\varepsilon\dfrac{s+2\lambda}{(s-\lambda)^2+\omega^2}$ | $\dfrac{\sigma_0}{\mu}\dfrac{s^{-1}+2\lambda s^{-2}}{(s+\lambda)^2+\omega^2}$ |
| Fractional Maxwell | $\dfrac{\Sigma_0}{\mu}\dfrac{1+\tau^{-\beta}s^{-\beta}}{s^2+\tau^{-\beta}s^{2-\beta}+\omega_\infty^2}$ | $\Delta\varepsilon\dfrac{s+\tau^{-\beta}s^{1-\beta}}{s^2+\tau^{-\beta}s^{2-\beta}+\omega_\infty^2}$ | $\dfrac{\sigma_0}{\mu}\dfrac{s^{-1}+\tau^{-\beta}s^{-1-\beta}}{s^2+\tau^{-\beta}s^{2-\beta}+\omega_\infty^2}$ |

TABLE IV. The cases of underdamping, overdamping, and critical damping in the Voigt model, corresponding to the classical damped oscillator. The notations for $\omega$ and $\lambda$ are detailed in the text.

| | |
|---|---|
| $\omega$ real | $\frac{\Sigma_0}{\mu\omega}\sin(\omega t)e^{-\lambda t}$ |
| $\omega$ imaginary | $\frac{\Sigma_0}{\mu|\omega|}\sinh(|\omega|t)e^{-\lambda t} \approx \frac{\Sigma_0}{2\mu\lambda}e^{-\frac{\omega_{eq}^2}{2\lambda}t}$ |
| $\omega = 0$ | $\frac{\Sigma_0}{\mu}t e^{-\lambda t}$ |

Figure 3 shows how the oscillatory behavior changes with the index $\beta$, reminding that at small fraction $\beta$ the damping element is close to an elastic unit in its response (hence the significant oscillations are retained). In this plot, the parameters are chosen such that $\omega_\infty = 1/2\tau$, corresponding to the critical damping regime. Hence when $\beta$ is close to one, the relaxation to the equilibrium strain is fast. The power-law determines the long time behavior and the equilibrium strain, with the decaying oscillations happening about this value. At short times, all solutions for a given $\beta$ converge, as all start with the same velocity and have the same spring constant, $G_r$. The system differs from the classical case in two key ways, first it has a long time memory of the impact owing to the power-law creep, whereas in the classical case the system is at rest as soon as the oscillations decay. The second difference is the system will always oscillate and there is no critical damping or overdamped situation in the strict sense, a consequence of there being no purely real solutions to $s^2 + \tau^{-1}s^{2-\beta} + \omega_\infty^2 = 0$ when $0 < \beta < 1$. However, there are still values of $\omega_\infty^2$ and $\tau$ which give the fastest decay of oscillations; this is discussed in more detail in Sec. V.

### B. Response to strain impulse

This regime practically is analogous to plucking a string or a suspended mass, and is perhaps less common in standard rheological testing. The inverse Laplace transformation of the Voigt model solution illustrated in Table III is given by the standard decaying oscillation

$$\frac{\varepsilon(t)}{\Delta\varepsilon} = \frac{\omega_{eq}}{\omega}\cos(\omega t + \phi)\exp(-\lambda t), \qquad (9)$$

where $\phi = \cos^{-1}(\omega/\omega_{eq})$, and as before: $\lambda = 2\eta/\mu$, $\omega_{eq}^2 = G_e/\mu$, and $\omega^2 = \omega_{eq}^2 - \lambda^2$. Here, we will once again have the underdamped, overdamped, and critical damped regimes when $\omega$ is real, imaginary, and zero, respectively. The strain rate can be simplified to a compact expression

$$\frac{1}{\Delta\varepsilon}\frac{d\varepsilon(t)}{dt} = -\frac{\omega_{eq}^2}{\omega}\sin[\omega t]\exp(-\lambda t). \qquad (10)$$

There is a subtle point here that the rate is not exactly out of phase with the strain, which is a direct consequence of the damping; as $\lambda \to 0$, then $\cos^{-1}\phi \to 0$ and they become properly 90° out of phase.

For the classical Maxwell model, the strain response is again

$$\frac{\varepsilon(t)}{\Delta\varepsilon} = \frac{\omega_\infty}{\omega}\cos(\omega t + \phi)\exp(-\lambda t). \qquad (11)$$

Here, $\phi = \cos^{-1}(\omega/\omega_\infty)$, and $\lambda = 1/2\tau$, $\omega_\infty^2 = G_r/\mu$, and $\omega^2 = \omega_\infty^2 - \lambda^2$. The expressions have the same form as in the Voigt model, although the $\lambda$ term has a different origin. It is clear why the Voigt model settles to zero strain in equilibrium, as there will always be a restoring force with the spring and dashpot in parallel. In this specific case of strain impulse, the dashpot of the Maxwell model does not extend in response to the initial instant deformation, which is taken by the spring; the system will then oscillate about zero strain, dissipating energy evenly above and below zero strain, eventually settling at zero strain.

The solution of the fractional Maxwell model is

$$\frac{\varepsilon(t)}{\Delta\varepsilon} = A\cos[\omega t + \phi]\exp(-\lambda t)$$
$$- \frac{\sin[\pi\beta]}{\pi}\int_0^\infty \rho(x)\exp(-xt/\tau)dx, \qquad (12)$$

where the detailed derivation, solutions for $\omega$ and $\lambda$, and the expressions for parameters $A$, $\phi$, and $\rho(x)$ are given in Appendix D. As with the classical Maxwell model, the solution returns to zero strain at equilibrium for the same fundamental reason of specific response to the strain impulse. In this case, the integral term has to decay slowly enough that it forces the strain remain below $\Delta\varepsilon$ at early times, countering the larger than one amplitude of the oscillations. However, as in the earlier case of stress impulse, this term must decay more quickly than oscillations themselves, leaving the decaying oscillations the only relevant response at longer times. Figure 4 shows that at $\beta$ close to one, that is, for an almost Newtonian dashpot, the response is close to the critical damping. Whereas at small $\beta$ the fractional element is closer to an elastic unit by its nature, and the response is only a weakly decaying oscillation.

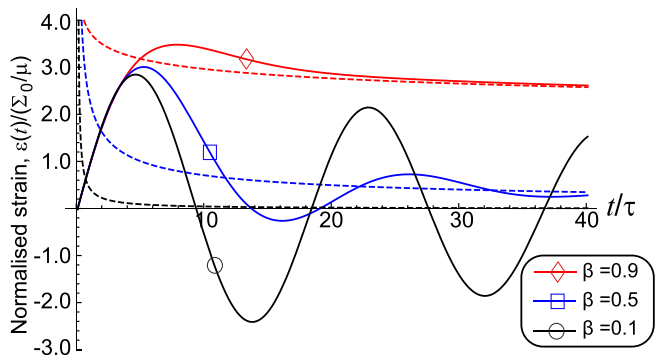

FIG. 3. The normalized response to a stress impulse, plotted as a function of scaled time $t/\tau$, for $\beta = 0.1$, $\beta = 0.5$, and $\beta = 0.9$. Parameters are chosen such that $\omega_\infty = 1/2\tau$, corresponding to the critical damping condition for the classical Maxwell model. The dashed lines show $\varepsilon(t) \propto t^{\beta-1}$, which becomes the equilibrium strain at long times. Note that all systems follow the same linear-growth response to an impact at $t < \tau$.

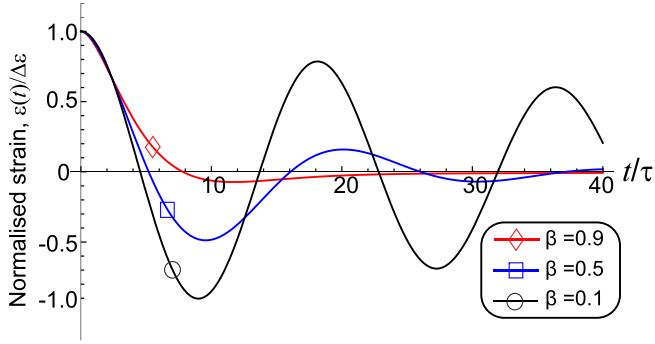

**FIG. 4.** The normalized response to a strain impulse as a function of scaled time, $t/\tau$, for $\beta = 0.1$, $\beta = 0.5$, and $\beta = 0.9$. The parameters are chosen such that $\omega_\infty = 1/2\tau$, corresponding to the critical damping regime of the classical Maxwell model.

## C. Response to step stress

This represents one of the common rheological experiments or situations, often called the creep compliance test. Taking the straightforward inverse Laplace transformation of the Voigt model solution for this initial condition (Table III), we obtain the decaying oscillation with respect to the equilibrium elastic deformation

$$\varepsilon(t) = \left(\frac{\sigma_0}{G_e}\right)\left(1 - \frac{\omega_{eq}}{\omega}\cos(\omega t + \phi)\exp(-\lambda t)\right), \quad (13)$$

where the parameters are the same as in the earlier cases of the Voigt model.

The solution to the classical Maxwell model reflects the equilibrium creep under constant stress

$$\varepsilon(t) = \dot{\varepsilon}_r t + \varepsilon_r\left(1 - \frac{1}{\omega_\infty^2\tau^2}\right) - \frac{\varepsilon_r\omega_\infty}{\omega}\cos[\omega t + \phi]\,\mathrm{e}^{-t/2\tau}. \quad (14)$$

Again, the parameters are the same as in the earlier examples of Maxwell model. The characteristic relaxation time after the onset of stress is $\tau = \eta/G_r$; $\varepsilon_r = \sigma_0/G_r$ is the extension of the spring element in the Maxwell model, and $\dot{\varepsilon}_r = \sigma_0/\eta$ is the extension rate of the dashpot element in this creep experiment. As we might expect, the transient regime has damped oscillations on top of the standard creep behavior of the Maxwell model.

Now examining the fractional case of the step stress response of a material, whose intrinsic relaxation is described by the Mittag-Leffler function, the solution for the time-dependent deformation is

$$\varepsilon(t) = \varepsilon_r\left(\frac{t^\beta}{\tau^\beta\Gamma[1+\beta]} - \frac{1}{\omega_\infty^2\tau^2}\frac{t^{2(\beta-1)}}{\tau^{2(\beta-1)}\Gamma[2\beta-1]} + 1 \right.$$
$$\left. - A\cos[\omega t + \phi]\mathrm{e}^{-\lambda t} - \frac{\sin[\pi\beta]}{\pi}\int_0^\infty \rho(x)\mathrm{e}^{-xt/\tau}\mathrm{d}x\right).$$
$$(15)$$

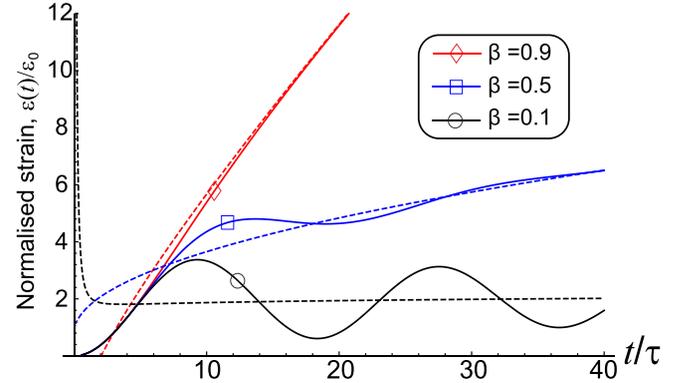

**FIG. 5.** The normalized response of the fractional Maxwell model to the step stress, plotted as a function of scaled time, $t/\tau$, for $\beta = 0.1$, $\beta = 0.5$, and $\beta = 0.9$. As before, we choose $\omega_\infty = 1/2\tau$, corresponding to critical damping in the classical Maxwell model. The dashed lines show the combined power-law terms in Eq. (15), settling for the equilibrium creep at long times. Note that, as in Figs. 3 and 4, the response at short times ($t < \tau$) is the same for all systems, in this case a parabolic constant-acceleration profile.

As usual, the details of derivation, expressions for $\omega$ and $\lambda$, as well as $A$, $\phi$, and $\rho(x)$ are given in the corresponding Appendix E. As in Eq. (8), we evaluate the integral containing $\rho(x)$ numerically. To address a possibility of poor convergence, we isolate this integral part to play as small a role as possible, bringing out the explicit long-time contributions which can be solved exactly.

Using the analogy with the classical case, we denote $\varepsilon_r = \sigma_0/G_r$ to be the constant contribution to strain from the extension of the spring under constant stress. Figure 5 illustrates the role of the fractional dashpot. Rather than linear (Newtonian) creep, we now find the power-law growth of deformation under the constant stress, following the $t^\beta$ law at long times. The second power-law term in Eq. (15), proportional to $t^{2(\beta-1)}$, represents a transient decay (since $\beta < 1$). This term, together with the integral term of $\rho(x)$, controls the initial regime and account for the time it takes for the mass to accelerate until the baseline creep reaches the equilibrium regime. As with the classical case, this is a more significant effect when the mass is large and the viscosity is low, as it takes a long time to accelerate to the equilibrium velocity; however, this is only relevant when $\beta$ is close to one. The response of a material with low $\beta$ is closer to a purely elastic element, with oscillations taking a very long time to decay. This type of oscillatory behavior is commonly seen in creep experiments where the instrument inertia cannot be avoided, and has been observed in materials with fractional Maxwell behavior [13,15], where they fit the oscillations with a numerical model.

## IV. RESPONSE TO DRIVEN OSCILLATIONS

In this section, we look at the response of our model viscoelastic materials to driven oscillations. As is typical in this case, we will focus on the equilibrium response, rather than transient solutions. At equilibrium, the solution for the amplitude of oscillations driven by an external stress $\sigma_0\sin(\omega t)$ takes the form

$$\varepsilon(t) = \frac{\sigma_0}{\mu\omega_\infty^2} A \sin[\omega t + \phi] = \varepsilon_r A \sin[\omega t + \phi], \qquad (16)$$

with

$$A = \sqrt{\frac{1 + (x\gamma)^{2\beta} + 2(x\gamma)^\beta \cos\left(\frac{\pi\beta}{2}\right)}{x^4 + (x\gamma)^{2\beta}(x^2-1)^2 + 2(x^2-1)(x\gamma)^\beta x^2 \cos\left(\frac{\pi\beta}{2}\right)}},$$

$$\phi = \tan^{-1}\left(\frac{(x\gamma)^\beta \sin\left(\frac{\pi\beta}{2}\right)}{x^2 + (x^2-1)x^{2\beta} + (x\gamma)^\beta(2x^2-1)\cos\left(\frac{\pi\beta}{2}\right)}\right).$$

As in other sections of this paper, we use notations: $x = \omega/\omega_\infty$ and $\gamma = \omega_\infty\tau = \eta/\sqrt{G_r\mu}$. Figure 6 shows the normalized strain oscillation amplitude, $A$, as a function of scaled frequency, $\omega/\omega_\infty$ for different values of fraction $\beta$. Plots (a) and (b) show the result for $\gamma = 0.01$ and $\gamma = 100$, respectively. One could think of low $\gamma$ meaning low viscosity, but high mass and/or relaxation modulus; high $\gamma$ represents low inertia spring and high viscosity situation. However, such interpretation is only strictly valid when $\beta = 1$.

For small $\gamma$, the dashpot is relatively weak and extends very easily, and as the spring and fractional dashpot are in series this means the spring plays a minor role in the oscillations. When $\beta$ is close to one, with lower frequency the amplitude gets much larger as the dashpot can flow much further for a given stress in with a longer period of oscillation. This is essentially the power law response of a fluid. Consequently, as frequency tends to zero the amplitude will tend to infinity, just as we would expect for constant stress on the classical Maxwell model [see Eq. (14)]. This is different in the Voigt model, or any model with a spring in parallel to the dashpot, where the amplitude would tend to the value of the spring with no damping elements. As $\beta$ gets lower, a resonance peak starts to rise, as both the value $\gamma^\beta$ gets larger and the dashpot becomes a weak solid contributing to the elasticity, which also allows the spring to play a greater role as the magnitudes become comparable. As we have a quasispring and an ordinary spring in series, their elastic contributions will also add in series, so the effective spring constant will be lower than the contribution of the weakest element, which in this case is the fractional dashpot due to the small $\gamma$. As $\beta$ decreases and $\gamma^\beta$ increases, the dashpot becomes more solid and the resonant frequency will increase, converging when $\beta = 0$. At $\beta = 0$, the fractional dashpot becomes a spring with modulus $G_r$, and the resonant frequency becomes $\omega_\infty/\sqrt{2}$ and the amplitude is double that of a single spring on its own.

In the case when $\gamma$ is large and $\beta$ is close to one, the dashpot will play relatively less role when compared to the spring and we see a typical spring resonance curve centered at $\omega_\infty$. However, for very low frequencies, the amplitude still becomes large as even though the dashpot has a high viscosity—it is still free to extend given enough time. As $\beta$ decreases, so will $\gamma^\beta$, and the fractional dashpot will become more comparable to the spring element, that is, contributing elastically. So we see both a lowering of the resonance peak due to greater movement of dashpot and higher damping, as well as a lowering of the resonant frequency. Once again, as $\beta$ tends to zero the fractional dashpot will become a spring of modulus $G_r$ and the resonant frequency is $\omega_\infty/\sqrt{2}$.

The shift of resonant frequency with $\beta$ is shown in greater detail in Fig. 7. Here, we plot the resonance frequency (defined as a position of the peak amplitude) as a continuous function of $\beta$ for several values of $\gamma$. In all cases, as $\beta$ tends to zero, the resonant frequency becomes $\omega_\infty/\sqrt{2}$, here $\gamma^\beta = 1$ and we left with two springs of the same modulus in series, irrespective of the value of viscosity $\eta$. For $\gamma$ close to, or below the critical damping condition ($\gamma = 1/2$), there is no resonant peak for large $\beta$, that is, when the viscous element in series is sufficiently small (perhaps counter-intuitively, in the Maxwell model higher viscosity means less damping, due to the series assembly). The value of $\beta$ needed to produce a resonant peak becomes smaller with $\gamma$, as the fractional dashpot needs to become more elastic before the spring in series can start storing energy, rather than energy simply being dissipating through the fractional dashpot.

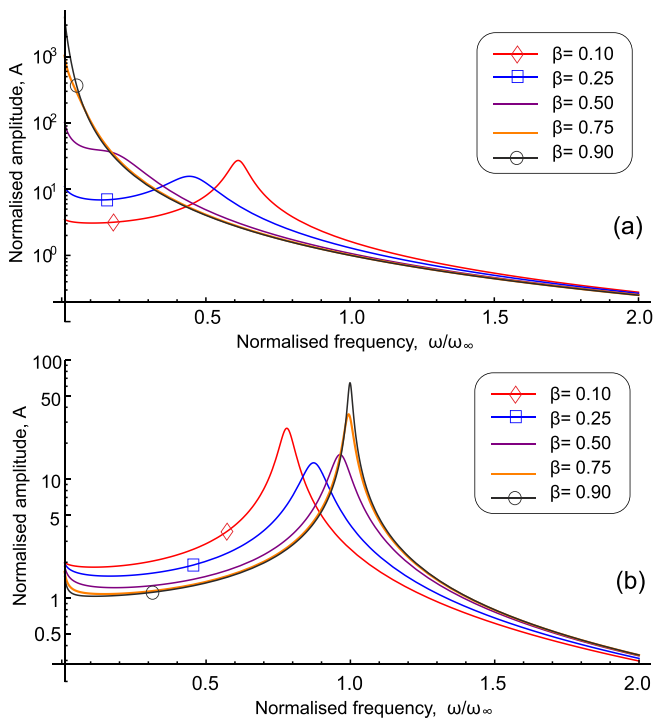

FIG. 6. The normalized amplitude of driven oscillation, $A$, plotted as a function of the scaled frequency, $\omega/\omega_\infty$, for several values of $\beta$ labeled in the graph: (a) for $\gamma = 0.01$, and (b) for $\gamma = 100$.

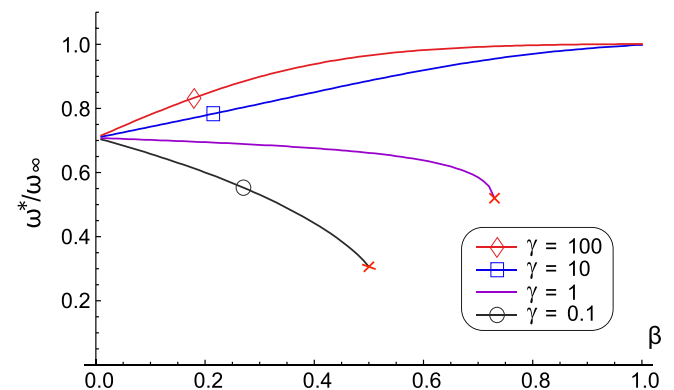

FIG. 7. The resonance frequency, $\omega^*/\omega_\infty$, plotted as a function of $\beta$ for four values of $\gamma = \omega_\infty\tau$. The red crosses indicate where the resonance peaks cease to exist.

## V. CRITICAL DAMPING

In contrast to the case of steady driven oscillation above, in this section we discuss the frequency and decay rate of oscillations during the transient regime. We specifically focus on how much damping occurs in different viscoelastic models in comparable conditions. All initial conditions for the imposed stress give essentially the same transient response in terms of oscillation frequency and damping rate. What differs between each case is the amplitude and phase difference of the oscillations, as well as the equilibrium strain. The oscillation frequency and the damping rate are determined by the location of the poles on the complex plane of the Laplace transformation (i.e., the solutions of $s^2 + \tau^{-\beta}s^{2-\beta} + \omega_\infty^2 = 0$): There are always two solutions, which are complex conjugates of each other. The real part, $-\lambda$, determines the decay rate of oscillations, while the imaginary part of these solutions, $\omega$, determines the frequency of the transient oscillations (see Appendix C for details). Figure 8 shows the plots of the dimensionless decay rate, $|\lambda|/\omega_\infty$, and the dimensionless frequency, $\omega/\omega_\infty$, as a function of the dimensionless variable $\gamma = \omega_\infty \tau$. We have seen in Sec. IV that $\gamma$ essentially measures the relative "strength" of the dissipative element compared to the inertial spring element of the mechanical model.

The two plots in Fig. 8 show how, for a given natural frequency $\omega_\infty = \sqrt{G_r/\mu}$, the decay rate and oscillation frequency change with changing the relaxation time $\tau$ for a given $\beta$. Remember that in terms of the fractional Maxwell model, $\tau = (b/G_r)^{1/\beta}$ (see Appendix A for details). For the classical Maxwell model, when $\beta = 1$, the peak decay rate is reached exactly at $\gamma = 1/2$ (i.e., the viscosity $2\eta = \sqrt{G_r\mu}$). The value of the peak decay rate is $\lambda = \omega_\infty$, meaning that transient oscillations decay over the half-period of the natural frequency. For $\gamma < 1/2$, the oscillation frequency is zero, and there are no oscillations with an increasingly long steady decay rate (the classically overdamped regime). Conversely, for $\gamma > 1/2$, the oscillation frequency rapidly converges to $\omega_\infty$ with an increasingly longer decay rate, corresponding to the classically underdamped regime.

However, these intuitive benchmarks become more obscure in the fractional relaxation case. On decreasing $\beta$, the decay rate has a much less defined peak, which becomes lower but also more spread out. This means that even though the maximum decay rate is lower, the significant damping occurs over a broad range of parameters (e.g., for systems with a different effective mass). As with the classical Maxwell case, the oscillation frequency increases with increasing $\gamma$; however, with $\beta \neq 1$ there is no classical

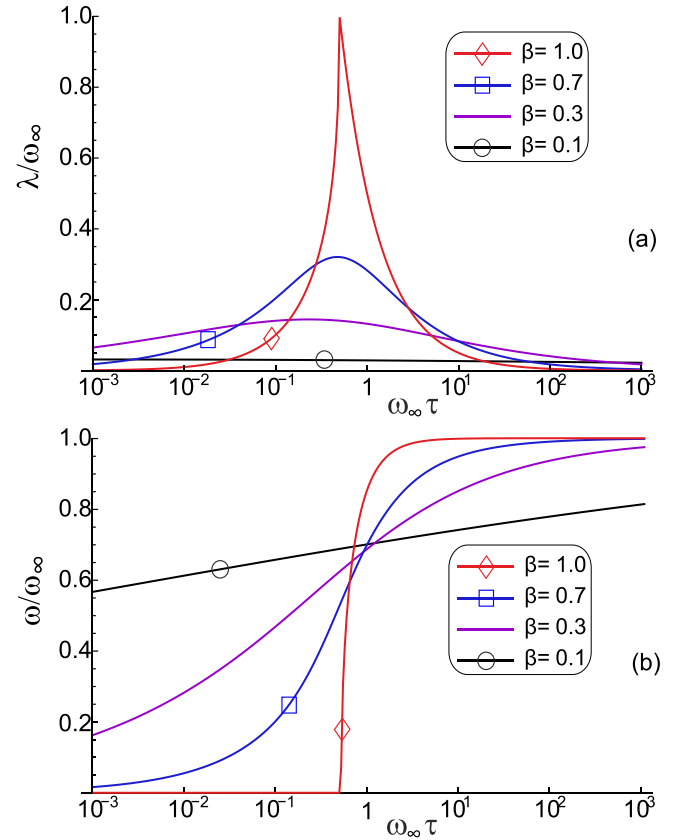

FIG. 8. The normalized decay constant, $|\lambda|/\omega_\infty$, plot (a), and the normalized frequency of transient oscillations, $\omega/\omega_\infty$, plot (b), as functions of dimensionless variable $\gamma = \omega_\infty \tau$ for several values of $\beta$ labeled in the graphs. Classically, the regime $\gamma < 1/2$ is overdamped, while at $\gamma > 1/2$ one finds underdamped oscillations; fractional nature of complex viscoelasticity makes this distinction blurred.

overdamped region—there always is a residual oscillation in a fractional viscoelastic system.

Figure 9 shows the critical damping factor $\gamma^* = \tau^*\omega_\infty$, at which there is a peak in the largest decay rate $|\lambda|$ in Fig. 8(a). In the classical case, this point also corresponds to the sharp transition between the overdamped and underdamped regimes. As $\beta$ decreases, the damping peak moves to lower $\gamma$. In the limiting case below $\beta \approx 0.1$ the peak is at very low values of $\gamma$, with a simple scaling relationship of $\gamma^* \simeq (1/2)^{1/\beta}$. In general, for low $\beta$ the peak value of damping becomes irrelevant, as the system acts primarily elastically with light damping that changes little over a broad range of $\omega\tau$.

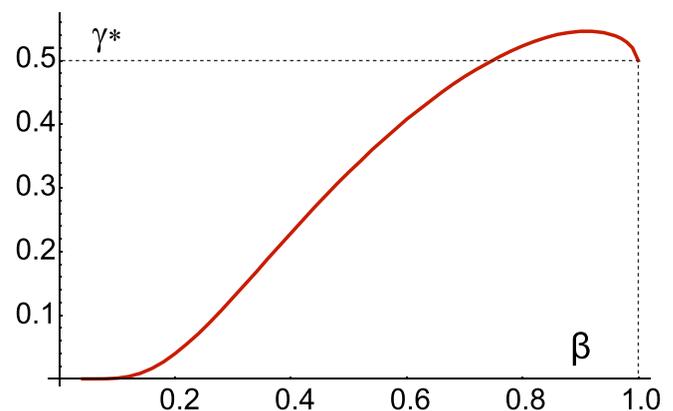

FIG. 9. The value of $\gamma$ at which the "critical damping" occurs: $\gamma^* = \tau^*\omega_\infty$, i.e., the position of peaks in Fig. 8(a), plotted as a function of continuously changing fraction $\beta$. Note that at low $\beta$ the damping peaks become so diffuse that the whole notion of critical damping becomes ill defined.

The interesting behavior occurs for the intermediate range of fractionality, $0.1 < \beta < 0.9$, where we still have significant damping but over a broader range of parameters. A viscoelastic material following the classical Maxwell relaxation behavior would be quite limited in terms of damping oscillations, as for a given $\tau$ and $G_r$, the reduced mass, $\mu$, would have to be tuned very carefully to give significant damping (the peak around the critical damping point is very sharp). In the fractional Maxwell model, for a given $\tau$, $G_r$, and $\beta$, the requirements on reduced mass are much less stringent for significant damping of oscillations to be present. This argument could be tuned to any other parameter from the relevant set, e.g., for a range of $G_r$ values for a fixed mass. This behavior is useful where the reduced mass (i.e., mass, area, and initial length of the construction) can change in the process, but damping is still required at a similar level.

A good example would be in the case of biological connective tissues, such as ligaments, tendons, or fascia. Each of these tissues has their own specific role and associated viscoelastic properties. For instance, tendon exhibits a low-exponent power-law relaxation behavior [37], and as such fits well with fractional Maxwell with a low $\beta < 0.05$. This agrees with the perceived biological function of a tendon, which is to passively store elastic energy allowing the muscle and soft tissues to dissipate energy at their own preferred rate without over-stretching [38,39]. It would be detrimental for a tendon function to have $\beta$ much greater than 0.1: It needs to act as an almost purely elastic element for a whole range of parameters (reduced mass, modulus, viscosity, rate of deformation), and that would be difficult to achieve in a construct made of inherently dissipative soft-matter materials. But by making $\beta$ very low, it can remain in this universally nondamping mode.

In contrast, soft tissues (such as fascia) have relaxation functions with higher values of $\beta \sim 0.2 - 0.4$, meaning that they can damp significantly over a relatively broad range of reduced mass, without the need to directly alter their mechanical properties. This is useful in running or in response to impact, where oscillations after a stress impulse must be rapidly dissipated in soft tissue (e.g., before the next stride) to avoid resonance and reinforcement of oscillation amplitude [40]. Similarly, natural rubber typically has a power-law decay of relaxation function with $\beta \sim 0.6$, and not surprisingly it is widely used in broad impact or vibration damping situations (from earthquake protection of buildings and resonance-proofing of bridges to vibration-insulation of cooling fans on circuit boards).

## VI. CONCLUSIONS

Fractional viscoelasticity is a relatively established field, driven by the practical need to describe materials with complex rheological response in a universal manner, using as few fitting parameters as possible [3,25]. We were especially motivated by the work of Rossikhin and Shitikova [18,19], who pioneered many steps that we had to follow in this work. Specifically, they have correctly solved the problem of stress impulse [obtaining the transfer function in Eq. (8) and Appendix C]. We reproduced their solution here with a particular focus on underlying physical processes, separating oscillating and nonoscillating parts, and highlighting relevant regimes. Several other types of rheological experiment (initial conditions) studied here, and the analysis of driven oscillation and critical damping is new in this field. Each of these problems has relevance in a different experimental setting, and these were discussed in the respective sections above.

In this paper, we focused on one specific aspect of complex viscoelasticity, which involves inertial effects and damping; in particular, it was having qualitative and significant effects in damping of oscillations. There are several other areas where the fractional calculus (and specifically—Maxwell model) plays a big role, for instance in the problems of anomalous diffusion in such media [41,42] and in associated approach to ageing [43].

One may ask why we have paid relatively less attention to the versions of Kelvin–Voigt model, which many would associate with a basic viscoelastic solid (as opposed to the classical Maxwell model that inherently shows plastic flow at long times). First of all, such intuitive understanding of Maxwell model representing viscoelastic creep is no longer valid once $\beta < 1$: We have seen how the fractional viscoelastic element effectively represents the internal elasticity of the system. There is also a subtle and little known issue with Voigt model in the fractional case, as both Naber [20] and Rossikhin [18] have noticed: Once a fractional damping element is inserted, there is a very abrupt transition in behavior (between $\beta = 1$ and $\beta = 0.99$). In the nominally overdamped regime, one expects [and indeed finds in the Maxwell model, cf. Fig. 8(a)] the rate of decay to be slow. However, as soon as the small fractionality is added to the Voigt model, the rate of decay becomes increasingly faster as one increases the magnitude of $\eta$ relative to $\sqrt{G_r\mu}$. At the same time, the fractional Voigt model is entirely valid and well-behaved (and thus corresponds to relevant experiments) in the underdamped regime. Our main point is that in fractional-relaxation systems, the Maxwell model is adequate for viscoelastic solids and predicts physically correct effects in oscillation and energy damping.

## ACKNOWLEDGMENTS

The authors are grateful to Fanlong Meng and Samuel Bell for many useful discussions. This research was supported by the EPSRC Critical Mass Grant for Cambridge Theoretical Condensed Matter (EP/J017639).

## APPENDIX A: RELAXATION FUNCTION OF THE FRACTIONAL MAXWELL MODEL

The relationship between stress and strain in the fractional Maxwell model arises from the series combination of the elastic element governed by $\sigma_1 = G_r\varepsilon_1$, and the fractional dashpot element governed by $\sigma_2 = b \mathrm{d}^\beta \varepsilon_2 / \mathrm{d}t^\beta$. As the elements are in series $\varepsilon = \varepsilon_1 + \varepsilon_2$ and $\sigma = \sigma_1 = \sigma_2$ in equilibrium, giving the stress strain relationship of

$$\sigma(t) + \tau^\beta \frac{\mathrm{d}^\beta \sigma(t)}{\mathrm{d}t^\beta} = \tau^\beta G_r \frac{\mathrm{d}^\beta \varepsilon(t)}{\mathrm{d}t^\beta}, \tag{A1}$$

where $\tau^\beta = b/G_r$. The dimensionality of the parameter $b$ must be a function of index $\beta$ for the units to balance. The relaxation function is the stress response to a step strain, and it can be shown either through Laplace transform or directly through the analysis of fractional derivatives [25] that this is given by

$$G(t) = G_r \mathrm{E}_\beta[-(t/\tau)^\beta], \tag{A2}$$

where $\tau = (b/G_r)^{1/\beta}$ is the characteristic relaxation time scale and $\mathrm{E}_\beta[z]$ is the Mittag-Leffler function named in honor of the Swedish mathematician Gösta M. Mittag-Leffler. The function is defined as the series expansion, which is convergent for the whole of the complex plane

$$\mathrm{E}_\beta[z] := \sum_{n=0}^{\infty} \frac{z^n}{\Gamma(\beta n + 1)}. \tag{A3}$$

This is a generalization of the basic exponential function, since when $\beta = 1$ one recovers $\Gamma(n + 1) = n!$ For a more detailed discussion, see Appendix E in the textbook [25].

## APPENDIX B: IMPACT IN A MAXWELL MODEL: INVERSE LAPLACE SOLUTION

"Impact" is the condition that we referred to as stress impulse above. Starting from the equation of motion in Laplace space we first rearrange it to isolate singularities in denominator. The first step is

$$\bar{\varepsilon}(s) = \frac{\Sigma_0}{\mu} \frac{1 + 2\lambda s^{-1}}{(s + \lambda)^2 + \omega^2}$$

$$\equiv \frac{\Sigma_0}{\mu} \left( \frac{1}{\omega} \frac{\omega}{(s + \lambda)^2 + \omega^2} + \frac{2\lambda}{s\left((s + \lambda)^2 + \omega^2\right)} \right). \tag{B1}$$

The second term can be split further

$$\frac{1}{s\left((s + \lambda)^2 + \omega^2\right)} \equiv \frac{1}{(\omega^2 - \lambda^2)s} + \frac{1}{\omega^2 - \lambda^2} \frac{-s - 2\lambda}{(s + \lambda)^2 + \omega^2}$$

$$\equiv \frac{1}{\omega_\infty^2} \left( \frac{1}{s} - \frac{s + \lambda}{(s + \lambda)^2 + \omega^2} \right.$$

$$\left. - \frac{\lambda}{\omega} \frac{\omega}{(s + \lambda)^2 + \omega^2} \right). \tag{B2}$$

Finally, we obtain the equation in a form where we can easily look up standard Laplace transforms

$$\bar{\varepsilon}(s) = \frac{2\lambda \Sigma_0}{\mu \omega_\infty^2} \left( \frac{1}{s} - \frac{s + \lambda}{(s + \lambda)^2 + \omega^2} \right.$$

$$\left. - \frac{\lambda}{\omega} \left( 1 - \frac{\omega_\infty^2}{2\lambda^2} \right) \frac{\omega}{(s + \lambda)^2 + \omega^2} \right). \tag{B3}$$

The three terms above correspond to the following parts in the equation of motion in time:

$$\varepsilon(t) = \frac{2\lambda \Sigma_0}{\mu \omega_\infty^2} \left( 1 - \cos(\omega t)\exp(-\lambda t) \right.$$

$$\left. - \frac{\lambda}{\omega} \left( 1 - \frac{\omega_\infty^2}{2\lambda^2} \right) \sin(\omega t)\exp(-\lambda t) \right), \tag{B4}$$

which can be assembled in compact form by joining the oscillating functions

$$\varepsilon(t) = \frac{2\lambda \Sigma_0}{\mu \omega_\infty^2} \left( 1 - \frac{\omega_\infty^2}{2\lambda \omega} \cos(\omega t + \phi)\exp(-\lambda t) \right), \tag{B5}$$

where $\phi = \cos^{-1}(2\lambda \omega / \omega_\infty^2)$.

## APPENDIX C: IMPACT IN A FRACTIONAL MAXWELL MODEL: INVERSE LAPLACE SOLUTION

Here, we consider the stress impulse situation again. The equation of motion in Laplace space can be manipulated into a form with isolated dimensionless singularities in denominator

$$\bar{\varepsilon}(s) = \frac{\Sigma_0}{\mu} \frac{1 + \tau^{-\beta} s^{-\beta}}{s^2 + \tau^{-\beta} s^{2-\beta} + \omega_\infty^2}$$

$$= \frac{\Sigma_0}{\omega_\infty^2 \mu} \left\{ \frac{1}{z^\beta} - \frac{(1 + z^\beta)z^{2(1-\beta)} - \gamma^2}{z^2 + z^{2-\beta} + \gamma^2} \right\}, \tag{C1}$$

where $0 \leq \beta \leq 1$, $z = \tau s$ and $\gamma = \omega_\infty \tau$. The solution can then be written as the sum of two functions

$$\varepsilon(t) = \left( \frac{\Sigma_0}{\omega_\infty^2 \mu} \right) \{ f_1(t) - f_2(f) \}, \tag{C2}$$

each representing the inverse Laplace transform of a term in Eq. (C1). For both functions, the inverse Laplace transform is part of the contour integral shown in Fig. 10, which due to the Cauchy theorem must be equivalent to the sum of enclosed residues. The paths along the negative real axis are branch cuts which account for the fractional power in the functions. The transformation involves calculating the following contour integrals:

$$f_1(t) = \frac{1}{2\pi i} \int_{\gamma - i\infty}^{\gamma + i\infty} \tilde{f}_1(z)\exp(zt/\tau)\mathrm{d}(z/\tau)$$

$$= \frac{1}{\tau} \left( \sum \text{Residues} - \sum \text{Branch cuts} \right). \tag{C3}$$

It can be shown that the function $\tilde{f}_1(s)$ has no residues, so only the branch cuts contribute. The branch cuts are the two path integrals from $-\infty$ to $0$ and $0$ to $-\infty$, where we substitute in $z \to -z = z\exp(i\pi)$ for the first integral and $z \to -z = z\exp(-i\pi)$ for the second integral

$$
\begin{aligned}
f_1(t) &= \frac{1}{2\pi i}\int_0^\infty \tilde{f}_1(ze^{-i\pi})\exp(-zt/\tau)\mathrm{d}z \\
&\quad - \frac{1}{2\pi i}\int_0^\infty \tilde{f}_1(ze^{i\pi})\exp(-zt/\tau)\mathrm{d}z \\
&= \frac{1}{2\pi i}\int_0^\infty \left(\frac{e^{-zt/\tau}}{z^\beta e^{-i\pi\beta}} - \frac{e^{-zt/\tau}}{z^\beta e^{i\pi\beta}}\right)\mathrm{d}z \\
&= \frac{1}{2\pi i}\int_0^\infty \frac{e^{-zt/\tau}}{z^\beta}(e^{i\pi\beta} - e^{-i\pi\beta})\mathrm{d}z \\
&= \frac{\sin[\pi\beta]}{\pi\tau}\int_0^\infty z^{-\beta}e^{-zt/\tau}\mathrm{d}z = \frac{\sin[\pi\beta]t^{\beta-1}}{\pi\tau^\beta}\int_0^\infty x^{-\beta}e^x\mathrm{d}x \\
&= \frac{t^{\beta-1}}{\tau^\beta}\frac{\sin[\pi\beta]\Gamma[1-\beta]}{\pi} = \frac{t^{\beta-1}}{\tau^\beta\Gamma[\beta]},
\end{aligned}
\tag{C4}
$$

where we used the substitution $x = zt/\tau$, and the identities $\Gamma[\alpha] = \int_0^\infty x^{\alpha-1}e^{-x}\mathrm{d}x$ and $\pi/\sin[\alpha\pi] = \Gamma[1-\alpha]\Gamma[\alpha]$. It can be quickly checked that a Laplace transform of this result gives $1/z^\beta$.

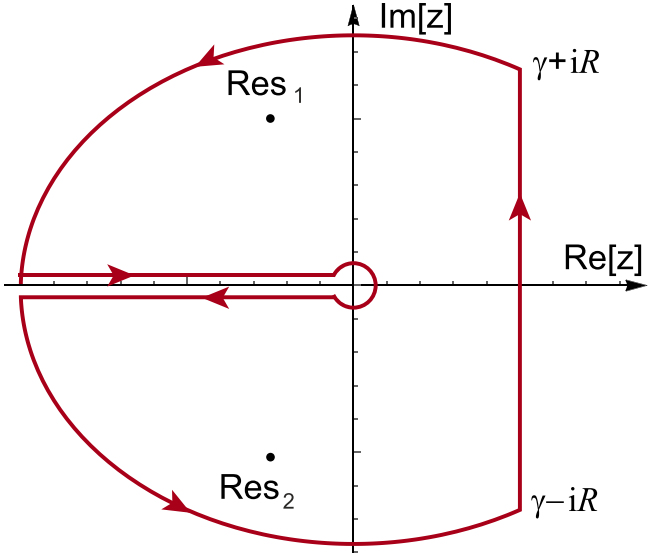

**FIG. 10.** The path for the contour integral of inverse Laplace transformation.

The second part of the inverse Laplace transform of $\tilde{f}_2(z)$ has complicated residues and branch cuts and requires careful analysis. As before, the branch cuts are given by substituting $z = -z = z\exp(i\pi)$ to the integral from $-\infty$ to $0$ and $z = -z = z\exp(-i\pi)$ to the integral from $0$ to $-\infty$ such that

$$
\begin{aligned}
\sum\text{Branchcuts} &= \frac{1}{2\pi i}\int_0^\infty \tilde{f}_2(ze^{i\pi})e^{-zt/\tau}\mathrm{d}(z/\tau) \\
&\quad - \frac{1}{2\pi i}\int_0^\infty \tilde{f}_2(ze^{-i\pi})e^{-zt/\tau}\mathrm{d}(z/\tau) \\
&= \frac{1}{2\pi i}\int_0^\infty \left[\frac{(1+z^\beta e^{i\pi\beta})z^{2(1-\beta)}e^{2i\pi(1-\beta)} - \gamma^2}{z^2 + z^{2-\beta}e^{i\pi(2-\beta)} + \gamma^2}\right. \\
&\quad \left. - \frac{(1+z^\beta e^{-i\pi\beta})z^{2(1-\beta)}e^{-2i\pi(1-\beta)} - \gamma^2}{z^2 + z^{2-\beta}e^{-i\pi(2-\beta)} + \gamma^2}\right]e^{-zt/\tau}\mathrm{d}z \\
&= -\frac{\sin[\pi\beta]}{\tau\pi}\int_0^\infty \rho(x)x^{-\beta}e^{-xt/\tau}\mathrm{d}x.
\end{aligned}
\tag{C5}
$$

We are left with a purely real integral over $x$, with a function in the integrand given by

$$
\rho(x) = \frac{x^{4-2\beta} + x^2(2\gamma^2 + x^2) + 2x^{2-\beta}(\gamma^2 + x^2)\cos[\pi\beta]}{\gamma^4 + x^{4-2\beta} + x^2(2\gamma^2 + x^2) + 2x^{2-\beta}(\gamma^2 + x^2)\cos[\pi\beta]}.
\tag{C6}
$$

Changing variables gives

$$
\sum\text{Branch cuts} = -\frac{\sin[\pi\beta]t^{\beta-1}}{\pi\tau^\beta}\int_0^\infty \rho\left(\frac{y\tau}{t}\right)y^{-\beta}e^{-y}\mathrm{d}y,
\tag{C7}
$$

multiplying the numerator and denominator of $\rho(\frac{y\tau}{t})$ by $t^4$. Note that as $t \to 0$, the function takes a limit $\rho \to 1$ and the branch cuts simplify to $-t^{\beta-1}/(\tau^\beta\Gamma[\beta])$ canceling the identical term in $f_1(t)$. This is a delicate but important point that prevents an unphysical divergence of inverse power-law solutions at $t \to 0$. As $t$ increases, the value of $\rho$ will decrease for any given $y$, and due to the $\exp[-y]$, this results in the integrand getting small rapidly. When $x > \gamma$ or $y > \omega_\infty\tau^2/t$, $\rho$ will once again tend to 1 owing to the $y^4$ term dominating over $\gamma$. However, by this time $\exp[-y]$ will be very small. In summary, the role of the branch cuts here is to counter the large value of $f_1(t)$ at early times, but then it quickly vanishes and has no influence on the equilibrium position or relevant regime of damped oscillations.

To calculate the residues, we must first find the location of the poles by solving $z^2 + z^{2-\beta} + \gamma^2 = 0$. This is most easily done by substituting in polar form $z = re^{i\theta}$ and separating into real and imaginary parts

$$
\begin{aligned}
r^2\cos[2\theta] + r^{2-\beta}\cos[(2-\beta)\theta] + \gamma^2 &= 0, \\
r^2\sin[2\theta] + r^{2-\beta}\sin[(2-\beta)\theta] &= 0.
\end{aligned}
\tag{C8}
$$

These equations can only be simultaneously be satisfied if the solution is to the left of the imaginary axis, and there are clearly two solutions which are complex conjugates of each other (i.e., at $\theta$ and $-\theta$). The pair of simultaneous equations, for $(r, \theta)$, are more usefully represented as

$$
\gamma^2 = \frac{\sin[\beta\theta]}{\sin[(2-\beta)\theta]}\left(-\frac{\sin[(2-\beta)\theta]}{\sin[2\theta]}\right)^{2/\beta}
\tag{C9}
$$

and

$$
r = \left(-\frac{\sin[(2-\beta)\theta]}{\sin[2\theta]}\right)^{1/\beta}.
\tag{C10}
$$

In the region of $\pi/2 < \theta < \pi$, $\sin[\beta\theta]$ is always positive and $\sin[2\theta]$ is always negative, hence Eq. (C9) is only satisfied if $\pi/2 < \theta < \pi/(2-\beta)$ where $\sin[(2-\beta)\theta]$ will be positive. The poles can be found to be first solving Eq. (C9) for the argument, $\theta$, for given values of $\beta$, $\tau$, and $\gamma$, this can then be substituted in Eq. (C10) to find the absolute value.

With the two poles ($z_1$ and $z_2$) known, the residues can now be calculated, as these are simple poles. The result is

$$\sum \text{Residues} = \lim_{z \to z_1}\left\{(z - z_1)\frac{(1 + z^\beta)z^{2(1-\beta)} - \gamma^2\, \mathrm{e}^{zt/\tau}}{z^2 + z^{2-\beta} + \gamma^2}\frac{\mathrm{e}^{zt/\tau}}{\tau}\right\} + \lim_{z \to z_2}\left\{(z - z_2)\frac{(1 + z^\beta)z^{2(1-\beta)} - \gamma^2\, \mathrm{e}^{zt/\tau}}{z^2 + z^{2-\beta} + \gamma^2}\frac{\mathrm{e}^{zt/\tau}}{\tau}\right\}. \quad (C11)$$

Taylor expanding each denominator about the pole and taking the limits leaves

$$\sum \text{Residues} = \frac{\gamma^2 - z_1^{2(1-\beta)}\left(1 + z_1^\beta\right)}{2z_1 + (2 - \beta)z_1^{1-\beta}}\frac{\mathrm{e}^{z_1 t/\tau}}{\tau} + \frac{\gamma^2 - z_2^{2(1-\beta)}\left(1 + z_2^\beta\right)}{2z_2 + (2 - \beta)z_2^{1-\beta}}\frac{\mathrm{e}^{z_2 t/\tau}}{\tau}$$

$$= \frac{\gamma^2 - z_1^{-\beta}\left(z_1^2 + z^{2-\beta}\right)}{2z_1 + (2 - \beta)z_1^{1-\beta}}\frac{\mathrm{e}^{z_1 t/\tau}}{\tau} + \frac{\gamma^2 - z_2^{-\beta}\left(z_2^2 + z_2^{2-\beta}\right)}{2z_2 + (2 - \beta)z_2^{1-\beta}}\frac{\mathrm{e}^{z_2 t/\tau}}{\tau}$$

$$= \gamma^2\frac{1 + z_1^{-\beta}}{2z_1 + (2 - \beta)z_1^{1-\beta}}\frac{\mathrm{e}^{z_1 t/\tau}}{\tau} + \gamma^2\frac{1 + z_2^{-\beta}}{2z_2 + (2 - \beta)z_2^{1-\beta}}\frac{\mathrm{e}^{z_2 t/\tau}}{\tau}$$

$$= \frac{\gamma^2}{\tau}\frac{z_1^\beta + 1}{2z_1^{\beta+1} + (2 - \beta)z_1}\mathrm{e}^{z_1 t/\tau} + \frac{\gamma^2}{\tau}\frac{z_2^\beta + 1}{2z_2^{\beta+1} + (2 - \beta)z_2}\mathrm{e}^{z_2 t/\tau}, \quad (C12)$$

where we have used the relation $z_{1,2}^2 + z_{1,2}^{2-\beta} + \gamma^2 = 0$. As $z_1 = z_2^*$, and $z_{1,2} = r\mathrm{e}^{\mathrm{i}\pi} = \tau(-\lambda + \mathrm{i}\omega)$, this can be simplified to

$$\sum \text{Residues} = A\cos[\omega t + \phi]\exp(-\lambda t). \quad (C13)$$

This gives the full solution as

$$\varepsilon(t) = \frac{\Sigma_0}{\mu\omega_\infty^2\tau}\left(\frac{t^{\beta-1}}{\tau^{\beta-1}\Gamma[\beta]} - \frac{\sin[\pi\beta]}{\pi}\int_0^\infty \rho(x)x^{-\beta}\exp(-xt/\tau)\mathrm{d}x - A\cos[\omega t + \phi]\exp(-\lambda t)\right), \quad (C14)$$

where the last term, containing most of the relevant information about the response, has the shorthand parameters

$$A = \gamma^2\left(\frac{1 + r^{2\beta} + 2r^\beta\cos[\beta\theta]}{r^{2\beta} + r^2(1 - \beta/2)^2 + r^{1+\beta}(2 - \beta)\cos[(1 - \beta)\theta]}\right)^{1/2} \quad (C15)$$

and

$$\phi = \tan^{-1}\left(-\frac{(2 - \beta)r\sin[\theta] - r^\beta(2\sin[\beta\theta] + (2 - \beta)r\sin[(1 - \beta)\theta])}{2r^{2\beta} + (2 - \beta)r\cos[\theta] + r^\beta(2\cos[\beta\theta] + (2 - \beta)r\cos[(1 - \beta)\theta])}\right). \quad (C16)$$

When $\beta = 1$, the decay integral vanishes, the power-law becomes a constant, and the solution is that of the classical Maxwell model. From examining the integrand, it becomes apparent that if $\gamma$ is small, then $\rho(x) \approx 1$: If this is the case, the integral will be equivalent to the power-law decay term, canceling it out and leaving only the oscillating part. For anything other than very small $\gamma$, the integral term will vanish relatively quickly. The power-law relaxation characterizes the long-term response of the system to the impulse, determining the equilibrium point about which any oscillations occur. It also implies that for $\beta < 1$, the system will eventually return to the original zero-strain value. That is, for any $\beta < 1$ the fractional Maxwell model retains a degree of elasticity, although this relaxation could take a very long time for larger $\beta$.

## APPENDIX D: STRAIN IMPULSE IN THE FRACTIONAL MAXWELL MODEL: INVERSE LAPLACE SOLUTION

The equation of motion for the Mittag-Leffler model in response to a strain impulse in Laplace space is given by

$$\frac{\tilde{\varepsilon}(s)}{\Delta\varepsilon} = \frac{s + \tau^{-\beta}s^{1-\beta}}{s^2 + \tau^{-\beta}s^{2-\beta} + \omega_\infty} \equiv \tau\frac{z + z^{1-\beta}}{z^2 + z^{2-\beta} + \gamma^2}, \quad (D1)$$

where $\omega_\infty^2 = G_r/\mu$, $z = s\tau$, and $\gamma = \omega_\infty\tau$. As the numerator is always finite for $0 \leq \beta \leq 1$, there are only two poles occurring when the denominator is zero. Then, as with [Appendix C](#), the solution is given by

$$\frac{1}{2\pi\mathrm{i}}\int_{\gamma-\mathrm{i}\infty}^{\gamma+\mathrm{i}\infty}\frac{\tilde{\varepsilon}(s)}{\Delta\varepsilon}\exp(st)\mathrm{d}s = \frac{1}{2\pi\tau\mathrm{i}}\int_{\gamma'\tau-\mathrm{i}\infty}^{\gamma'\tau+\mathrm{i}\infty}\frac{\tilde{\varepsilon}(z)}{\Delta\varepsilon}\exp(zt/\tau)\mathrm{d}z = \sum \text{Residues} - \sum \text{Branchcuts}. \quad (D2)$$

The branchcuts are given by

$$\sum \text{Branchcuts} = \frac{1}{2\pi\tau i} \int_0^\infty \left( \frac{\tilde{\varepsilon}(ze^{i\pi})}{\Delta\varepsilon} - \frac{\tilde{\varepsilon}(ze^{-i\pi})}{\Delta\varepsilon} \right) e^{-zt/\tau} dz = \frac{\sin[\pi\beta]}{\pi} \int_0^\infty \rho(x) e^{-xt/\tau} dx, \tag{D3}$$

where

$$\rho(x) = \frac{\gamma^2 x^{1-\beta}}{\gamma^4 + x^{4-2\beta} + x^2(2\gamma^2 + x^2) + 2x^{2-\beta}(\gamma^2 + x^2)\cos(\pi\beta)}. \tag{D4}$$

The poles are in the same location as those in Appendix C and determined by Eqs. (C9) and (C10). So,

$$\sum \text{Residues} = \frac{z_1 + z_1^{1-\beta}}{2z_1 + (2-\beta)z_1^{1-\beta}} e^{z_1 t/\tau} + \frac{z_2 + z_2^{1-\beta}}{2z_2 + (2-\beta)z_2^{1-\beta}} e^{z_2 t/\tau}, \tag{D5}$$

as $z_1 = z_2^*$, $z_1 = re^{i\pi} = \tau(-\lambda + i\omega)$, and $a\cos[\omega t] + b\sin[\omega t] = \sqrt{a^2 + b^2}\cos[\omega t + \tan^{-1}(-b/a)]$ this can be simplified to

$$\sum \text{Residues} = A\cos[\omega t + \phi]\exp(-\lambda t), \tag{D6}$$

where

$$A = \left( 1 - \frac{1 - (\beta/4) + r^\beta \cos[\beta\theta]}{1 + r^{2\beta} + 2r^\beta \cos[\beta\theta]} \right)^{-1/2}, \quad \text{and} \quad \phi = \tan^{-1}\left( -\frac{\beta r^\beta \sin[\beta\theta]}{2 - \beta + 2r^\beta + (4-\beta)r^\beta \cos[\beta\theta]} \right),$$

giving the full equation of motion as

$$\frac{\varepsilon(t)}{\Delta\varepsilon} = A\cos[\omega t + \phi]\exp(-\lambda t) - \frac{\sin[\pi\beta]}{\pi} \int_0^\infty \rho(x)\exp(-xt/\tau)dx. \tag{D7}$$

## APPENDIX E: STEP STRESS IN THE CLASSICAL AND FRACTIONAL MAXWELL MODELS: INVERSE LAPLACE SOLUTION

The solution to the Maxwell mode is given by

$$\begin{aligned}
\tilde{\varepsilon}(s) &= \left(\frac{\sigma_0}{\mu}\right) \frac{s^{-1}}{s^2 + s\tilde{G}(s)/\mu} = \left(\frac{\sigma_0}{\mu}\right) \frac{1 + 2\lambda s^{-1}}{s\left((s+\lambda)^2 + \omega^2\right)} \\
&= \frac{\sigma_0}{\mu\omega_\infty^2} \left( \frac{2\lambda}{s^2} + \left(1 - \frac{4\lambda^2}{\omega_\infty^2}\right)\left(\frac{1}{s} - \frac{s+\lambda}{(s+\lambda)^2 + \omega^2}\right) + \frac{\lambda}{\omega}\left(\frac{4\lambda^2}{\omega_\infty^2} - 3\right)\frac{\omega}{(s+\lambda)^2 + \omega^2} \right) \\
&= \frac{\dot{\varepsilon}_\infty}{s^2} + \varepsilon_\infty\left( \left(1 - \frac{\dot{\varepsilon}_\infty}{\tau\omega_\infty^2}\right)\left(\frac{1}{s} - \frac{s+\lambda}{(s+\lambda)^2 + \omega^2} - \frac{\lambda}{\omega}\frac{\omega}{(s+\lambda)^2 + \omega^2}\right) - \frac{2\lambda}{\omega}\frac{\omega}{(s+\lambda)^2 + \omega^2} \right) \\
&= \frac{\dot{\varepsilon}_r}{s^2} + \varepsilon_r\left( \left(1 - \frac{1}{\omega_\infty^2\tau^2}\right)\left(\frac{1}{s} - \frac{s+\lambda}{(s+\lambda)^2 + \omega^2}\right) - \frac{1}{\omega\tau}\left(3 - \frac{1}{\omega_\infty^2\tau^2}\right)\frac{\omega}{(s+\lambda)^2 + \omega^2} \right), \tag{E1}
\end{aligned}$$

where $\lambda = 1/2\tau$, $\omega^2 = \omega_\infty^2 - \lambda^2$, $\omega_\infty^2 = G_r/\mu$, and $\tau = \eta/G_r$. $\varepsilon_r = \sigma_0/G_r$ and is the extension of the spring element in the Maxwell model in a creep experiment, and $\dot{\varepsilon}_r = \sigma_0/\eta$ and is the extension rate of the dashpot element in a creep experiment. The inverse is given by

$$\begin{aligned}
\varepsilon(t) &= \dot{\varepsilon}_r t + \varepsilon_r\left( \left(1 - \frac{1}{\omega_\infty^2\tau^2}\right)\left(1 - \cos[\omega t]e^{-t/2\tau}\right) - \frac{1}{\omega\tau}\left(3 - \frac{1}{\omega_\infty^2\tau^2}\right)\sin[\omega t]e^{-t/2\tau} \right) \\
&= \dot{\varepsilon}_r t + \varepsilon_r\left(1 - \frac{1}{\omega_\infty^2\tau^2}\right) - \frac{\varepsilon_r\omega_\infty}{\omega}\cos[\omega t + \phi]e^{-t/2\tau}, \tag{E2}
\end{aligned}$$

where

$$\phi = \tan^{-1}\left[\frac{1}{2\tau\omega}\left(\frac{1 - \omega_\infty^2\tau^2}{1 + \omega_\infty^2\tau^2}\right)\right] = \cos^{-1}\left[\frac{\omega}{\omega_\infty}\left(1 - \frac{1}{\omega_\infty^2\tau^2}\right)\right]. \tag{E3}$$

The equation of motion for the fractional Maxwell model is given by the Laplace transform of Mittag-Leffler function. The response to a step stress in Laplace space is given by

$$\frac{\tilde{\varepsilon}(s)}{(\sigma_0/\mu)} = \frac{s^{-1-\beta}(s^\beta + \tau^{-\beta})}{s^2 + \tau^{-\beta}s^{2-\beta} + \omega_\infty} \equiv \tau^3 \frac{1 + z^\beta}{z^{1+\beta}(z^2 + z^{2-\beta} + \gamma^2)}, \tag{E4}$$

where $\omega_\infty^2 = G_r/\mu$, $z = s\tau$, and $\gamma = \omega_\infty\tau$. With further rearranging, we get

$$\begin{aligned}
\frac{\tilde{\varepsilon}(z)}{(\sigma_0/\tau^3\mu)} &= \frac{1}{z(z^2 + z^{2-\beta} + \gamma^2)} + \frac{1}{z^{1+\beta}(z^2 + z^{2-\beta} + \gamma^2)} \\
&= \frac{1}{\gamma^2}\left(\frac{1}{z} + \frac{1}{z^{1+\beta}} - \frac{z + 2z^{1-\beta} + z^{1-2\beta}}{z^2 + z^{2-\beta} + \gamma^2}\right) \\
&= \frac{1}{\gamma^2}\left(\frac{1}{z} + \frac{1}{z^{1+\beta}} - \frac{1}{\gamma^2}\frac{1}{z^{2\beta-1}} - \frac{1}{\gamma^2}\frac{(z + 2z^{1-\beta})\gamma^2 - z^{3-2\beta} - z^{3-3\beta}}{z^2 + z^{2-\beta} + \gamma^2}\right).
\end{aligned}$$

The inverse Laplace transformation is then composed of four distinct terms

$$\varepsilon(t) = \frac{\sigma_0\tau}{\omega_\infty^2\mu}\left\{f_1(t) + f_2(t) - f_3(t) - f_4(t)\right\}, \tag{E5}$$

with

$$f_1(t) = \mathcal{L}^{-1}\left\{\frac{1}{z}\right\}(t); \quad f_2(t) = \mathcal{L}^{-1}\left\{\frac{1}{z^{1+\beta}}\right\}(t); \quad f_3(t) = \mathcal{L}^{-1}\left\{\frac{1}{\gamma^2}\frac{1}{z^{2\beta-1}}\right\}(t); \quad f_4(t) = \mathcal{L}^{-1}\left\{\frac{1}{\gamma^2}\frac{(z + 2z^{1-\beta})\gamma^2 - z^{3-2\beta} - z^{3-3\beta}}{z^2 + z^{2-\beta} + \gamma^2}\right\}(t).$$

The first function, $f_1(t)$, is simply a constant

$$f_1(t) = \mathcal{L}^{-1}\left\{\frac{1}{z}\right\}(t) = \mathcal{L}^{-1}\left\{\frac{1}{s\tau}\right\} = \frac{1}{\tau}, \tag{E6}$$

and will correspond to the equilibrium strain in the spring part of the fractional Maxwell model. For $f_2$, the integral can be split into two parts, the first is the branch cuts and the second is small circular path integral about the origin. The circular path integral can be written as

$$\frac{1}{2\pi i}\oint_\gamma \frac{e^{zt}}{z^{1+\beta}}dz = \frac{1}{2\pi i}\int_\pi^{-\pi}\frac{e^{t\rho e^{i\theta}}}{\rho^{1+\beta}e^{i\theta(1+\beta)}}i\rho e^{i\theta}d\theta, \tag{E7}$$

this does not vanish as $\rho \to 0$ so there is a problem here. However, with analogy to Eq. (C4) we can trial the solution

$$f_2(t) = \frac{t^\beta}{\tau^{\beta+1}\Gamma[1 + \beta]}, \tag{E8}$$

and see if we recover the Laplace transform

$$\tilde{f}_2(s) = \int_0^\infty \frac{t^\beta}{\tau^{1+\beta}\Gamma[1+\beta]}e^{-st}dt = \int_0^\infty \frac{x^\beta}{s^{1+\beta}\tau^{1+\beta}\Gamma[1+\beta]}e^{-x}dx, \quad \tilde{f}_2(z) = \frac{1}{z^{1+\beta}}\frac{1}{\Gamma[\alpha]}\int_0^\infty x^{\alpha-1}e^{-x}dx = \frac{1}{z^{1+\beta}},$$

which is what we expected. By analogy, we can use a similar trial solution for $f_3(t)$

$$f_3(t) = \frac{1}{\gamma^2\tau^{2\beta-1}}\frac{t^{2\beta-2}}{\Gamma[2\beta - 1]}, \tag{E9}$$

which again we can check via the Laplace transform

$$\tilde{f}_3(s) = \int_0^\infty \frac{t^{2\beta-2}}{\gamma^2\tau^{2\beta-1}\Gamma[2\beta-1]}\mathrm{e}^{-st}\mathrm{d}t = \frac{s^{1-2\beta}}{\gamma^2\tau^{2\beta-1}\Gamma[2\beta-1]}\int_0^\infty x^{2\beta-2}\mathrm{e}^{-x}\mathrm{d}x, \quad \tilde{f}_3(z) = \frac{1}{\gamma^2 z^{2\beta-1}\Gamma[\alpha]}\int_0^1 x^{\alpha-1}\mathrm{e}^{-x}\mathrm{d}x = \frac{1}{\gamma^2 z^{2\beta-1}}.$$

Looking at $\tilde{f}_4(z)$, we have

$$
\begin{aligned}
\text{Branch cuts} &= \frac{1}{2\pi\mathrm{i}\tau\gamma^2}\left(\int_{-\infty}^0 \frac{(z+2z^{1-\beta})\gamma^2 - z^{3-2\beta} - z^{3-3\beta}}{z^2 + z^{2-\beta} + \gamma^2}\mathrm{e}^{zt/\tau}\mathrm{d}z + \int_0^{-\infty} \frac{(z+2z^{1-\beta})\gamma^2 - z^{3-2\beta} - z^{3-3\beta}}{z^2 + z^{2-\beta} + \gamma^2}\mathrm{e}^{zt/\tau}\mathrm{d}z\right) \\
&= \frac{1}{2\pi\mathrm{i}\tau\gamma^2}\left(\int_{-\infty}^0 \frac{(-z+2z^{1-\beta}\mathrm{e}^{\mathrm{i}\pi(1-\beta)})\gamma^2 - z^{3-2\beta}\mathrm{e}^{\mathrm{i}\pi(3-2\beta)} - z^{3-3\beta}\mathrm{e}^{\mathrm{i}\pi(3-3\beta)}}{z^2 + z^{2-\beta}\mathrm{e}^{\mathrm{i}\pi(2-\beta)} + \gamma^2}\mathrm{e}^{-zt/\tau}\mathrm{d}z\right. \\
&\quad \left. + \int_0^{-\infty} \frac{(-z+2z^{1-\beta}\mathrm{e}^{-\mathrm{i}\pi(1-\beta)})\gamma^2 - z^{3-2\beta}\mathrm{e}^{-\mathrm{i}\pi(3-2\beta)} - z^{3-3\beta}}{z^2 + z^{2-\beta}\mathrm{e}^{-\mathrm{i}\pi(2-\beta)} + \gamma^2}\mathrm{e}^{-zt/\tau}\mathrm{d}z\right) \\
&= \frac{\sin[\pi\beta]}{\pi\tau\gamma^2}\int_0^\infty \rho(x)\mathrm{e}^{-tx/\tau}\mathrm{d}x,
\end{aligned}
$$

where

$$\rho(x) = \frac{x^2\big(\gamma^2 + 2\cos[2\pi\beta]\big(\gamma^2+x^2\big) + 2x^2\big) + 2x^3\cos[\pi\beta]\big(\gamma^2 + x^{2-2\beta} + x^2\big) - \gamma^2 x^{\beta+1}\big(2\gamma^2 + x^2\big)}{x^4 + x^{2\beta}(\gamma^2+x^2)^2 + 2x^{\beta+2}(\gamma^2+x^2)\cos[\pi\beta]}, \tag{E10}$$

and as before made the change of variable from $z \to -z = z\mathrm{e}^{\mathrm{i}\pi}$ for the upper integral and $z \to -z = z\mathrm{e}^{-\mathrm{i}\pi}$ for the lower integral, which after rearranging results in a solely real integral.

Next, we need to look at the part contributed by the residues, as with the previous solution we have simple poles which are complex conjugates of each other

$$
\begin{aligned}
\sum\text{Residues} &= \frac{1}{\tau\gamma^2}\sum_{i=1,2}\frac{\big(z_i+2z_i^{1-\beta}\big)\gamma^2 - z_i^{3-2\beta} - z_i^{3-3\beta}}{2z_i + (2-\beta)z_i^{1-\beta}}\mathrm{e}^{z_it/\tau} = \frac{1}{\tau\gamma^2}\sum_{i=1,2}\frac{z_i^\beta\big(z_i^2 + z_i^{2-\beta} + z_i^{2-\beta}\big)\gamma^2 - z^{2-\beta}\big(z_i^2 + z^{2-\beta}\big)}{z^{1+\beta}\big(2z_i + (2-\beta)z_i^{1-\beta}\big)}\mathrm{e}^{z_it/\tau} \\
&= \frac{1}{\tau\gamma^2}\sum_{i=1,2}\frac{z_i^\beta\big(-\gamma^2 + z_i^{2-\beta}\big)\gamma^2 + z^{2-\beta}\gamma^2}{z^{1+\beta}\big(2z_i + (2-\beta)z_i^{1-\beta}\big)}\mathrm{e}^{z_it/\tau} = \frac{1}{\tau}\sum_{i=1,2}\frac{z_i^2 + z^{2-\beta} - \gamma^2 z_i^\beta}{z^{1+\beta}\big(2z_i + (2-\beta)z_i^{1-\beta}\big)}\mathrm{e}^{z_it/\tau} \\
&= -\frac{\gamma^2}{\tau}\sum_{i=1,2}\frac{1 + z_i^\beta}{z^{1+\beta}\big(2z_i + (2-\beta)z_i^{1-\beta}\big)}\mathrm{e}^{z_it/\tau}, \tag{E11}
\end{aligned}
$$

where we have used the relation $z_{1,2}^2 + z_{1,2}^{2-\beta} + \gamma^2 = 0$. As $z_1 = z_2^*$, $z_1 = r\mathrm{e}^{\mathrm{i}\pi} = \tau(-\lambda + \mathrm{i}\omega)$, and $a\cos[\omega t] + b\sin[\omega t] = \sqrt{a^2+b^2}\cos[\omega t + \tan^{-1}(-b/a)]$, this can be simplified to

$$\sum\text{Residues} = -\frac{\gamma^2}{\tau}A\cos[\omega t + \phi]\exp(-\lambda t), \tag{E12}$$

where

$$A = \frac{1}{r^2}\left(1 - \beta\frac{1 - (\beta/4) + r^\beta\cos[\beta\theta]}{1 + r^{2\beta} + 2r^\beta\cos[\beta\theta]}\right)^{-1/2};$$

and

$$\phi = \tan^{-1}\left(-\frac{(2+2r^{2\beta}-\beta)\sin[2\theta] + r^\beta(2-\beta)\sin[(2-\beta)\theta] + 2r^\beta\sin[(2+\beta)\theta]}{(2+2r^{2\beta}-\beta)\cos[2\theta] + r^\beta(2-\beta)\cos[(2-\beta)\theta] + 2r^\beta\cos[(2+\beta)\theta]}\right),$$

so we find that

$$f_4(t) = -\frac{\gamma^2}{\tau}A\cos[\omega t + \phi]\exp(-\lambda t) - \frac{\sin[\pi\beta]}{\pi\tau}\int_0^\infty \rho(x)\mathrm{e}^{-tx/\tau}\mathrm{d}x. \tag{E13}$$

Combining all the elements together has the full solution in the form

$$\varepsilon(t) = \frac{\sigma_0 \tau}{\omega_\infty^2 \mu} \left( \frac{1}{\tau} + \frac{t^\beta}{\tau^{\beta+1} \Gamma[1+\beta]} + \frac{1}{\gamma^2 \tau^{2\beta-1}} \frac{t^{2\beta-2}}{\Gamma[2\beta-1]} \right.$$
$$\left. + \frac{\gamma^2}{\tau} A \cos[\omega t + \phi] e^{-\lambda t} + \frac{\sin[\pi\beta]}{\pi\tau} \int_0^\infty \rho(x) e^{-tx/\tau} dx \right).$$
$$\text{(E14)}$$

## APPENDIX F: DRIVEN OSCILLATIONS

The general oscillating solution in response to the driving force $\sigma_0 \sin[\omega t]$ takes the form in Laplace space

$$\bar{\varepsilon}(s) = \frac{\sigma_0}{\mu} \frac{\omega}{(s^2 + \omega^2)} \frac{s^{-1}}{s^2 + s\bar{G}(s)/\mu}. \quad \text{(F1)}$$

With the application of fractional Maxwell model, this becomes

$$\bar{\varepsilon}(s) = \frac{\sigma_0}{\mu} \frac{\omega}{(s^2 + \omega^2)} \frac{1 + \tau^{-\beta} s^{-\beta}}{(s^2 + \tau^{-\beta} s^{2-\beta} + \omega_\infty^2)}, \quad \text{(F2)}$$

and using the substitutions of $z = s\tau$, $\gamma^2 = \tau^2 \omega_\infty^2$, and $\alpha = \omega\tau$, we get

$$\bar{\varepsilon}(z) = \frac{\sigma_0 \tau^3 \alpha}{\mu} \frac{1}{(z + i\alpha)(z - i\alpha)} \frac{1 + z^{-\beta}}{z^2 + z^{2-\beta} + \gamma^2}. \quad \text{(F3)}$$

We are interested in the equilibrium response rather than the transient response, which comes about from the poles caused by $z = \pm i\alpha$. The branch cuts will vanish as $t \to \infty$, as will the other two residues as these poles will always have a negative real part. So keeping only the parts that do not vanish at $t = \infty$, the solution becomes

$$\varepsilon(t \to \infty) = \frac{1}{\tau} \sum_{i=1,2} \lim_{z \to z_i} \left[ (z - z_i)\bar{\varepsilon}(z) e^{zt/\tau} \right]$$
$$= \frac{\sigma_0 \tau^2 \alpha}{\mu} \sum_{i=1,2} \frac{1}{2z_i} \frac{1 + z_i^{-\beta}}{z_i^2 + z_i^{2-\beta} + \gamma^2} e^{z_i/\tau}, \quad \text{(F4)}$$

As $z_1 = z_2^*$, $z_1 = e^{i\pi/2} = i\tau\omega$, and $a\cos[\omega t] + b\sin[\omega t] = \sqrt{a^2 + b^2} \sin[\omega t + \tan^{-1}(b/a)]$, this can be simplified to

$$\varepsilon(t) = \frac{\sigma_0}{\mu} A \sin[\omega t + \phi], \quad \text{(F5)}$$

where

$$A = \sqrt{\frac{1 + (x\gamma)^{2\beta} + 2(x\gamma)^\beta \cos\left(\frac{\pi\beta}{2}\right)}{x^4 + (x\gamma)^{2\beta}(x^2 - 1)^2 + 2(x^2 - 1)(x\gamma)^\beta x^2 \cos\left(\frac{\pi\beta}{2}\right)}};$$

and

$$\phi = \tan^{-1} \left( \frac{(x\gamma)^\beta \sin\left(\frac{\pi\beta}{2}\right)}{x^2 + (x^2 - 1)x^{2\beta} + (x\gamma)^\beta(2x^2 - 1)\cos\left(\frac{\pi\beta}{2}\right)} \right),$$

where $x = \omega/\omega_\infty$ and $\gamma = \omega_\infty \tau = \eta/\sqrt{G_r \mu}$.